\documentclass[prx,twocolumn,amsmath,amssymb]{revtex4}

\usepackage[dvipdfmx]{xcolor}
\usepackage{bm}
\usepackage{dcolumn}
\usepackage[pdftex]{graphicx}    

\usepackage{slashed}
\usepackage{amsmath}\usepackage{accents}
\usepackage[normalem]{ulem}
\usepackage{physics}

\newcommand{\iPi}{\Pi}

\newcommand{\cin}[1]{{\color{black}#1}}

\newcommand{\1}{\mbox{1}\hspace{-0.25em}\mbox{l}}

\begin{document}

\title{
Flow of unitary matrices:
Real-space winding numbers in one and three dimensions
}

\author{Fumina Hamano and Takahiro Fukui}
\affiliation{Department of Physics, Ibaraki University, Mito 310-8512, Japan}

\date{\today}

\begin{abstract}
The notion of the flow introduced by Kitaev is a manifestly topological formulation of the winding number on a real lattice.
First, we show in this paper that the flow is quite useful for practical numerical computations for systems without translational invariance. 
Second, we extend
it to three dimensions. Namely, we derive a formula of the flow on a three-dimensional lattice, which corresponds to 
the conventional winding number when systems have translational invariance.
\end{abstract}

\pacs{
}

\maketitle

\section{Introduction}

Topological classification of states in  condensed matter physics \cite{Zirnbauer:1996wi,Altland:1997aa,Schnyder:2008aa,Kitaev:2009fk,Teo:2010fk,Shiozaki:2014aa}
has been extended to various systems that are not necessarily so regular.
For example, topologically protected edge modes were reported in geophysics as well as biophysics \cite{Delplace:2017ui,yamauchi2020chiralitydriven}.
In real circumstances such systems are far from regular, 
which does not allow to calculate the Berry curvature as a function of the wave vectors.
Nevertheless, topological protection ensures robustness of the topological edge states.
Accordingly, direct computational schemes of topological invariants  for various irregular systems  
have become increasingly important.

For disordered and/or interacting systems, there are many attempts at computing topological invariants in real spaces.
One way is the use of the twisted boundary conditions, where twist angles play a role of the wave vectors \cite{Niu:1985fr}.
However, note that the integration of the Berry curvature over the twist angles has no clear physical reason:
Topological numbers should be basically attributed to fixed boundary conditions.
Indeed, in the case of the Chern number, if one uses the discretized plaquette method \cite{FHS05} for a sufficiently large system, 
just one plaquette, in principle, reproduces the correct Chern number \cite{PhysRevLett.122.146601}. 
The merit of this method is that it is always integral, and for clean noninteracting systems, it reduces to the topological 
invariants based on the Berry curvatures.
Another way is based on 
the direct real-space representations of topological invariants
\cite{Kitaev:2006yg,PhysRevB.80.125327,Prodan_2010,PhysRevB.84.241106,PhysRevB.89.224203,Caio:2019aa,PhysRevA.100.023610,PhysRevLett.122.166602,PhysRevA.101.063606,PhysRevB.103.224208,PhysRevB.103.155134,PhysRevA.103.043310}. 
The Zak phase is the typical example of them
\cite{Zak:1989fk}, which is also 
the basis of the quantum mechanical theory of electric polarization in crystalline insulators 
\cite{King-Smith:1993aa,Vanderbilt:1993fk}.
However, the topological nature of them seems unclear at first sight.

In this paper, we restrict our discussions to winding numbers of unitary matrices in odd dimensions, \cin{which 
are topological invariants characterizing half-filled ground states of systems with chiral symmetry.}
As mentioned above, real-space representations \cin{of topological invariants} are more or less based on the quantum mechanical 
position operators.
However, Kitaev \cite{Kitaev:2006yg} has proposed  a quite useful notion,  the {\it flow} of the unitary matrices.
This is equivalent to the Zak-like representation by the use of the position operators if the systems have translational symmetry.
Moreover, the flow in one dimension is manifestly topological. 
The purpose of the paper is firstly to show the usefulness of the flow also in the practical computations, 
and secondly to present  the three-dimensional extension of the flow. 
Recently, a method of computing a winding number in a discretized wave-vector space in three dimensions 
has been proposed in Ref. \cite{shiozaki2024discrete}. 
The three-dimensional flow in this paper is an alternative discrete formulation of the
three-dimensional winding number.

\section{Flow of unitary matrices in one dimension}\label{s:onedim}

\cin{In condensed matter physics, topological invariants are defined on the torus (the Brillouin zone) 
spanned by the continuum wave vector, which implies that corresponding real spaces are composed of infinite lattices.
Indeed, the flow introduced by Kitaev is basically defined by unitary matrices $U_{ij}$ specified by 
site indices $i,j$ running from $-\infty$ to $+\infty$. 
In other words, only for infinite dimensional matrices, the flow is topological. Practically, this feature is 
rather problematic, especially for numerical computations for finite size systems, since the flow vanishes trivially.

Keeping these points in mind,} we review in this section the flow of unitary matrices introduced by Kitaev \cite{Kitaev:2006yg}, using 
the Su-Schrieffer-Heeger (SSH) model \cite{Su:1979aa} as a typical example, \cin{stressing why the flow vanishes for finite systems, and how to overcome this difficulty for numerical calculations using finite systems.}

\cin{
Before proceeding, let us fix our notation of unitary matrices often denoted as $U_{ij}$. 
The indices $i,j$ stand for the lattice sites of unit cells, often referred to simply as sites, 
and if the systems have any $n$ internal degrees of freedom such as 
orbitals, $U_{ij}$ is a $n\times n$ matrix specified by $i,j$. 
The symbol $\rm tr$ means the trace over the internal degrees of freedom, whereas 
Tr stands for the trace including sites ${\rm Tr }\,U=\sum_i{\rm tr } \,U_{ii}$.
}

\subsection{SSH model}

Let us start with the generalized SSH model  described by the Hamiltonian on an infinite lattice:
\begin{alignat}1
\cin{\hat H}&
\cin{=(\bm c_A^\dagger,\bm c_B^\dagger)
\left(\begin{array}{cc}& \Delta\\ \Delta^\dagger &\end{array}\right)
\left(\begin{array}{c}\bm c_A\\ \bm c_B\end{array}\right)},
\label{SSHHam}
\end{alignat}
where $\bm c_{A}^\dagger=(\dots, c_{A,-1}^\dagger, c_{A,0}^\dagger, c_{A,1}^\dagger,c_{A,2}^\dagger,\dots)$, 
and likewise for $\bm c_{B}$, 
and the hopping matrix $\Delta$ is defined by
\begin{alignat}1
\Delta=\left(
\begin{array}{cccccccc}
\ddots &&&&&&&\\
&t_1&&&&&&\\
&t_2&t_1&&&&\\
&t_3&t_2&t_1&&&&\\
&&t_3&t_2&t_1&&&\\
&&&t_3&t_2&t_1&&\\
&&&&t_3&t_2&t_1&\\
&&&&&&&\ddots
\end{array}
\right).
\label{Delinf}
\end{alignat}
\cin{The basic symmetry of the Hamiltonian Eq. (\ref{SSHHam}) is chiral symmetry. 
In addition, the model possesses time-reversal symmetry, so that the model belongs to class BDI 
\cite{Zirnbauer:1996wi,Altland:1997aa,Schnyder:2008aa,Kitaev:2009fk}.
The half-filled ground state is topologically characterized by the winding number 
of the Fourier-transformed matrix $\Delta$.}
Note that $t_3$ is a specific hopping for a nontrivial high winding number \cite{PhysRevB.89.224203,PhysRevB.103.224208}. 
For the bulk system without disorder,  the Fourier transformation gives $\Delta=t_1+t_2e^{ik}+t_3e^{2ik}$.
Now, let us try to calculate the winding number in the lattice space, using the flow of Kitaev.
To this end, let us unitarize the matrix $\Delta$ by using the singular value decomposition $\Delta=VGW^\dagger$ such that
\begin{alignat}1
U=VW^\dagger. 
\label{UniTar}
\end{alignat}
We separate the one-dimensional lattice sites \cin{specifying the positions of the unit cells as follows:
Let $j$ be the label of the unit cell. Then, let us separate them} into two regions, say, $j\ge0$ and $j<0$, and let us call them 
region $A=1$ and 2, respectively.
Now, according to Kitaev \cite{Kitaev:2006yg}, we introduce  the flow of $U$ as
\begin{alignat}1
{\cal F}_1(U)
&=\sum_{j\ge0,k<0}\cin{\tr}\left(U^\dagger_{kj}U_{jk}-U^\dagger_{jk}U_{kj}\right),
\label{Flo}
\end{alignat}
\cin{where in the present SSH model, tr is needless.}
Let us define the projector onto region 1 as $\Pi_1\equiv \Pi$,  
where $\Pi_{ij}=\delta_{ij}$ for $i\ge0$ and $\Pi_{ij}=0$ otherwise, and
the projector onto region 2 as $\Pi_2\equiv \1-\Pi$. Then,  
the flow can be written as
\begin{alignat}1
{\cal F}_1(U)&=\Tr\left(\epsilon^{AB}U^\dagger \Pi_A U\Pi_B\right)
=\Tr \left(\cin{U^\dagger}\Pi U-\Pi\right)
\nonumber\\
&=\Tr U^\dagger[\Pi, U].
\label{DefFlo}
\end{alignat}
When a system has translational invariance, this reduces to Eqs. (\ref{TraInv_1}) and (\ref{TraInv_2}), 
\begin{alignat}1
{\cal F}_1(U)={\cal W}_1(U)\equiv \frac{i}{2\pi}\int dq \,\tr \,U_q^\dagger \partial_q U_q,
\label{WinNum_1}
\end{alignat}
where ${\cal W}_1(U)$ is the conventional winding number of the unitary matrix $U$ defined in the Fourier space.
In the present case of the SSH model, $U_q$ is just a single complex number, so that the trace in Eq. (\ref{WinNum_1}) is
not necessary.
Without translational symmetry, the winding number ${\cal W}_1$ cannot be defined, whereas the flow ${\cal F}_1$ is well defined.

For an infinite system, two matrices inside Tr in Eq. (\ref{DefFlo}) are infinite dimensional, so that 
the Tr operation should be carried out after the subtraction of the two matrices.
It should be noted that $U^\dagger \Pi U$ is a projector having 0 or 1 eigenvalues.
Thus,  ${\cal F}_1$ is integer valued. Moreover, it is manifestly topological, since even if a \cin{unit cell} 
in region 1 is assigned to region 2, \cin{i.e., even if the regions 1 and 2 are deformed,} 
the flow is invariant as Kitaev showed \cite{Kitaev:2006yg}. \cin{(See also Appendix \ref{s:topo}.)}
While the flow counts the difference of eigenvalue 1 between $U^\dagger \Pi U$ and $\Pi$, we can give 
an alternative formulation which counts the difference of 0's,  as presented in Appendix \ref{s:index}.
This may be a kind of the index theorem.

As proposed by Kitaev,  the above trace can be evaluated as if it were for finite dimensional matrices, when one 
introduces the truncation projector $\Pi^{(L)}$
\begin{alignat}1
{\cal F}_1&=\Tr \iPi^{(L)} U^\dagger\iPi U-\Tr\Pi^{(L)} \iPi,
\label{DefFlo_2}
\end{alignat}
where 
\begin{alignat}1
\iPi^{(L)}_{ij}=\left\{\begin{array}{ll}\delta_{ij} & (-L\le i\le L-1)\\0&(\mbox{otherwise})\end{array}\right..
\label{TruOpe}
\end{alignat}
This formula may be useful for practical numerical applications, although it spoils the integer nature of ${\cal F}_1$.
Below, let us show some examples.

\subsubsection{Topological phase}
The topological phase of the conventional SSH model ($t_3=0$) is adiabatically deformed to the model with 
$t_1=0$ and $t_2=1$ \cite{Ryu:2002fk}. In this case, $\Delta$ is already a unitary matrix without using the singular value decomposition Eq. 
 (\ref{UniTar}),  and we find \cin{for $U=\Delta$,}
\begin{alignat}1
U^\dagger \iPi U&=
\left(
\begin{array}{cccc|cccc}
\ddots &&&&&&&\\
&0&&&&&&\\
&&0&&&&&\\
&&&1&&&&\\
\hline
&&&&1&&&\\
&&&&&1&&\\
&&&&&&1&\\
&&&&&&&\ddots
\end{array}
\right)
\nonumber\\
&\equiv
\mbox{ diag} \left(\dots,0,0,1|1,1,1,\dots\right),
\nonumber\\
\Pi& =
\mbox{ diag} \left(\dots,0,0,0|1,1,1,\dots\right),
\label{BulFlo}
\end{alignat}
where the straight lines in the matrix stand for the boundaries separated by $\Pi_{1,2}$.
Thus, we have ${\cal F}_1=1$. 
\cin{This is very sharp contrast to the trivial phase below in Sec. \ref{s:bulktrivial}. 
One knows that $\Delta$ in this case actually moves a particle
to its neighbor. Although nothing  can  be found in $U^\dagger U=\1$, the boundary introduced by the projector $\Pi$ 
in between $U^\dagger$ and $U$ reveals the 
flow just at the boundary, as can be seen in Eq. (\ref{BulFlo}). It would be a kind of the bulk-edge correspondence.}

\subsubsection{Trivial phase} \label{s:bulktrivial}

Let us set $t_2=0$ and $t_1=1$. Then, $\Delta=\1$ is a unitary matrix. In this case, $U^\dagger \iPi U=\Pi$.
Thus, we have ${\cal F}_1=0$.

\subsection{Finite systems with the periodic boundary condition}\label{s:onedimf}
\label{s:finite}

For numerical calculations, we inevitably use finite systems. 
With the periodic boundary condition, $\Delta$ in Eq. (\ref{Delinf}) becomes
\begin{alignat}1
\Delta=
\left(
\begin{array}{ccccccc}
t_1&&&&&t_3&t_2\\
t_2&t_1&&&&&t_3\\
t_3&t_2&t_1&&&&\\
&&&\ddots&&&\\
&&&t_2&t_1&&\\
&&&t_3&t_2&t_1&\\
&&&&t_3&t_2&t_1
\end{array}
\right). 
\label{DelFin}
\end{alignat}
Let us assume that the number of the unit cells of the SSH model is finite, $2N$.
The matrix $\Delta$ is then $2N\times 2N$ matrix. 
Let us separate the sites into two sets $0\le j\le N-1$ and $-N\le j\le -1$,  called region 1 and 2, respectively,
and introduce corresponding projectors $\Pi_1\equiv\Pi$ and $\Pi_2\equiv\1-\Pi$, similarly in the infinite system.

\subsubsection{Topological phase}

As in the case of the infinite lattice, the matrix $U$ with winding number 1 is adiabatically deformed  into
\begin{alignat}1
U=
\left(
\begin{array}{ccccccc}
0&&&&&&1\\
1&0&&&&&\\
&1&0&&&&\\
&&&\ddots&&&\\
&&&&0&&\\
&&&&1&0&\\
&&&&&1&0
\end{array}
\right). \quad
\end{alignat}
The top right matrix element $U_{-N,N-1}=1$ is due to the periodic boundary condition.
This matrix $U$ is unitary, and we have
\begin{alignat}1
U^\dagger\iPi U&=
\mbox{diag}\left(0,0,\dots,0,1|1,1,\dots,1,0\right).
\label{NonTriFin}
\end{alignat}
The finite-size effect is manifest as the matrix element $(U^\dagger\iPi U)_{N-1,N-1}=0$ (the last 0 in the above):
For the infinite one-dimentional (1D) chain, the projector $\Pi$ chooses the space $j\ge0$ which has only one boundary at $j=0$,
whereas in the periodic chain, the projector gives rise to two boundaries. 
\cin{Since the flow occurs at one direction, a positive flow at one boundary induces a negative flow at the other boundary,
implying vanishing total flow.}
Therefore, we have ${\cal F}_1=0$ for the finite-size system even for the SSH model in the topological phase. 
This is also expected from 
the conventional identity associated with the trace, $\Tr U^\dagger \Pi U=\Tr UU^\dagger\Pi=\Tr\Pi$ holds 
for finite-dimensional matrices.
Nevertheless,
 the truncation projector $\Pi^{(L)}$ in Eq. (\ref{TruOpe}) is useful, in practice, even for finite systems.
Namely, we have for Eq. (\ref{NonTriFin})
\begin{alignat}1
{\cal F}_1=\Tr \iPi^{(L)}(U^\dagger\Pi U-\Pi)=1,
\label{FloUni}
\end{alignat}
where $\Pi^{(L)}$ is defined by Eq. (\ref{TruOpe}), if one chooses $L$ as $1\le L\le N-1$.
\cin{This truncation projector removes the flow at an artificial boundary due to finite-size effects.}
\begin{figure}[h]
\begin{center}
\begin{tabular}{c}
\includegraphics[width=0.99\linewidth]{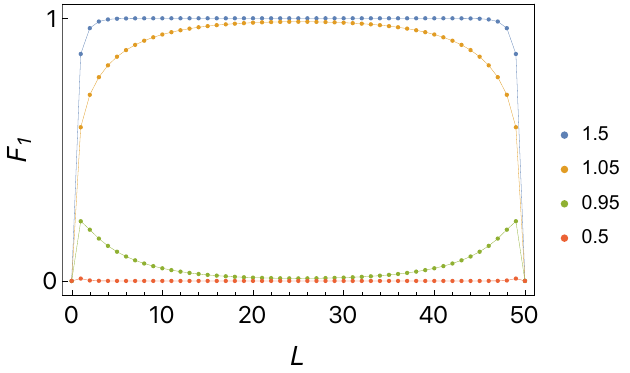}
\end{tabular}
\caption{
The flow Eq. (\ref{FloUni})
as a function of the truncation size $L$ for the finite unitary matrix Eq. (\ref{DelFin}) under the periodic boundary condition
with $t_2=1.5,1.05,0.95,0.5$. The other parameters  $t_1=1$ and $t_3=0$ and the system size $2N=100$ are fixed. 
${\cal F}_1=0$ at $L=N$ reflects the fact that finite-size systems always show the trivial flow without the truncation.
}
\label{f:flow}
\end{center}
\end{figure}

In Fig. \ref{f:flow}, we show the flow as a function of the truncation size $L$ near the SSH transition point $t_1=t_2$ $(t_3=0)$. 
It turns out that the flow does not so strongly 
depend on $L$, and the size $L\sim N/2$ may be suitable to reproduce the correct topological transition.

\subsection{Application to the SSH model with disorder}

\begin{figure}[h]
\begin{center}
\begin{tabular}{c}
\includegraphics[width=0.99\linewidth]{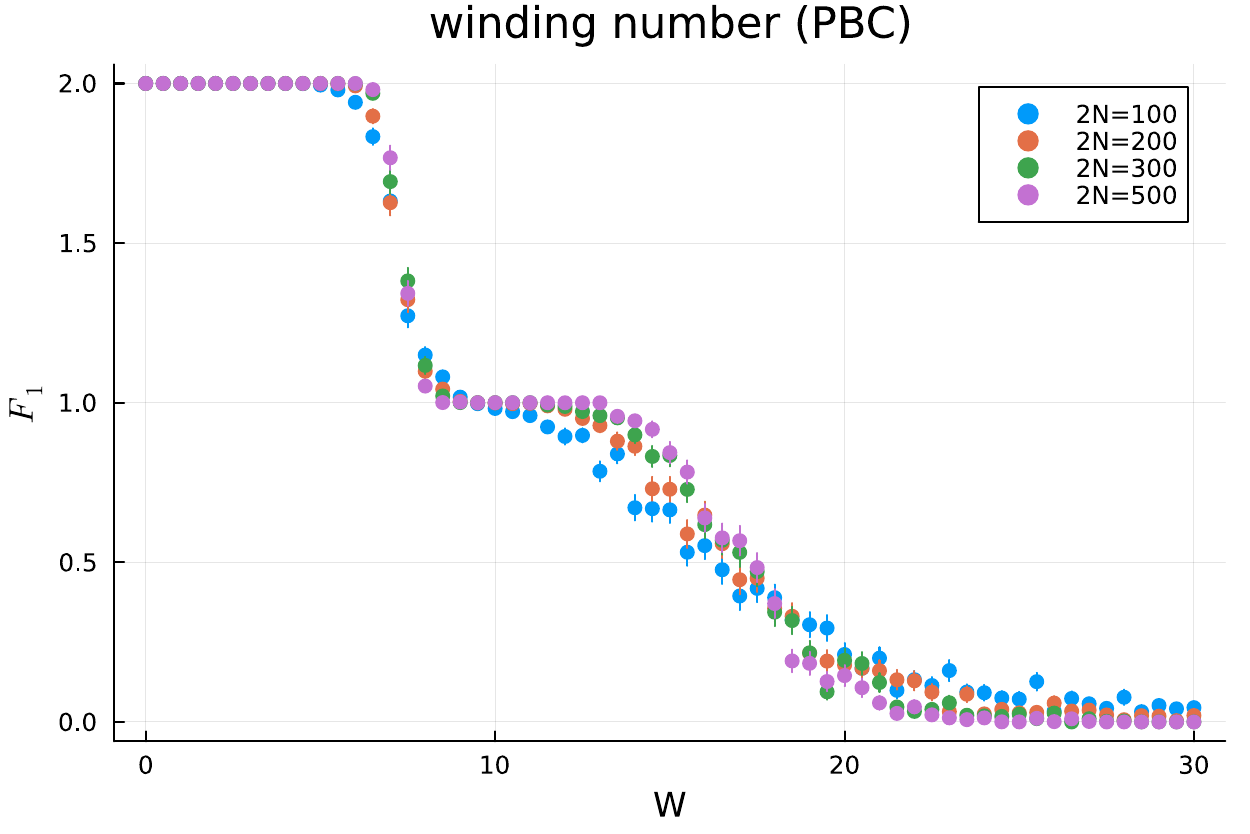}
\end{tabular}
\caption{
The flow as a function of the disorder strength $W$ of the generalized SSH model 
with size $2N=100$, 200, 300 and 500 averaged over 100 ensembles.
}
\label{f:bench}
\end{center}
\end{figure}

As an example of the calculation of the flow, we examine the generalized SSH model studied in 
Refs. \cite{PhysRevB.89.224203,PhysRevB.103.224208}, which gives topological transition due to disorder.
We set
\begin{alignat}1
t_{1,j}&=0+\delta t_{1,j},
\nonumber\\
t_{2,j}&=1+\delta t_{2,j},
\nonumber\\
t_3&=-2 ,
\end{alignat}
where $\delta t_{1,j}\in[-W/2,W/2]$ and $\delta t_{2,j}\in[-W/4,W/4]$ are random parameters.
In Fig. \ref{f:bench}, we show the flow ${\cal F}_1$ as a function of the disorder strength for the generalized SSH model. 
It turns out that the flow can reveal the topological transitions quantitatively, and thus the truncation scheme works around 
the transition points of dirty systems.
\cin{As discussed in Ref. \cite{PhysRevB.89.224203}, the sequential topological transition ${\cal F}_1=2\rightarrow1\rightarrow0$ shows 
that the present model can be characterized by the winding numbers rather than the polarizations (the Berry phase) with $Z_2$ nature.
Although it may be difficult to find the model in the real materials, the metamaterial such as topoelectrical circuits could give an experimental platform for observing the sequential topological transitions.}

\section{Flow in three dimensions}

The flow introduced by Kitaev is basically for one dimension, 
but the same idea leads to the Chern number in two dimensions, which is
represented by the projectors to the occupied states on a lattice \cite{Kitaev:2006yg}. 
In this section, we further generalize the flow to three dimensions, and derive the real-space winding number on
a three dimensional lattice.

\subsection{Winding number}

Let $U_q$ be a unitary matrix as a function of the wave vector $q=(q_x,q_y,q_z)$. Then, 
the winding number of  $U_q$  is defined by
\begin{alignat}1
{\cal W}_3(U)=\frac{1}{24\pi^2}\int d^3q \epsilon^{\mu\nu\rho}
\tr(U^\dagger\partial_\mu UU^\dagger\partial_\nu UU^\dagger\partial_\rho U),
\label{WinNum_3}
\end{alignat}
where $\partial_\mu\equiv\partial_{q_\mu}$.
On a lattice, the site dependence of the matrix $U$ is generically denoted as $U_{jk}$, 
where $j=(j_x,j_y,j_z)$ and $k$ label two sites. 
In the present case with translational invariance, we assume $U_{jk}=U_{j-k}$.
Then, $U_{j-k}$ is related with $U_q$ via the Fourier transformation
\begin{alignat}1
U_{j-k}=\int_{-\pi}^\pi \frac{d^3q}{(2\pi)^3}e^{iq\cdot (j-k)}U_q.
\end{alignat} 
Using Eq. (\ref{TraInv_2}), the winding number Eq. (\ref{WinNum_3}) can be rewritten in the real space as
\begin{alignat}1
{\cal W}_3(U)&=\frac{\pi i}{3}\tr\left(\epsilon^{\mu\nu\rho}U^\dagger[X_\mu, U]U^\dagger[X_\nu, U]U^\dagger[X_\rho, U]\right),
\label{3DWinNum_pos}
\end{alignat}
where $X_{\mu,ij}=i_\mu\delta_{ij}$ is the position operator.
Equation (\ref{TraInv_1}) further leads to 
\begin{alignat}1
{\cal W}_3(U)
&=\frac{\pi i}{3}\Tr\left(\epsilon^{\mu\nu\rho}U^\dagger[\Pi_\mu, U]U^\dagger[\Pi_\nu, U]U^\dagger[\Pi_\rho, U]\right)
\nonumber\\
&=-\frac{\pi i}{3}\Tr \left(\epsilon^{\mu\nu\rho}U^\dagger\Pi_\mu U \Pi_\nu U^\dagger \Pi_\rho U\right)
\nonumber\\
&\equiv -\frac{\pi i}{3}w_3(U,\Pi_x,\Pi_y,\Pi_z),
\label{3DWinNum}
\end{alignat}
where $\Pi_\mu$ is the projector onto the non-negative $\mu$ direction, 
$(\Pi_\mu)_{jk}=\delta_{jk}$ for $j_\mu\ge 0$ and $=0$ otherwise.

\subsection{Flow}\label{s:3Dflow}

Let us separate the three-dimensional lattice spanned by $j=(j_x,j_y,j_z)$ into 
four regions denoted by $A=1,2,3,4$  and introduce corresponding 
projectors $\Pi_A$. For example, we can choose each region such that
$j_x\ge0,j_y\ge0,j_z<0$ ($A=1$), $j_x<0, j_y\ge0,j_z<0$ ($A=2$), $j_z\ge0$ ($A=3$),
and $j_y<0,j_z<0$ ($A=4$).
We assume $\Pi_A\Pi_B=\delta_{AB}\Pi_A$  and $\sum_A\Pi_A=\1$
with $A=1,2,3,4$, for simplicity.
Define the flow by
\begin{alignat}1
{\cal F}_3(U)&=-2\pi i\Tr \left(\epsilon^{ABCD}U^\dagger \Pi_A U\Pi_B U^\dagger \Pi_C U\Pi_D\right).
\label{DefFlo3}
\end{alignat}
This may be a generalized definition of the flow Eq. (\ref{DefFlo}) to three dimensions.
Using $\Pi_4=1-(\Pi_1+\Pi_2+\Pi_3)$, we have
\begin{alignat}1
{\cal F}_3(U)&=-2\pi i\Tr \left(\epsilon^{ABC}U^\dagger \Pi_A U\Pi_B U^\dagger \Pi_C U\right),
\nonumber\\
& \equiv -2\pi if_3(U,\Pi_1,\Pi_2,\Pi_3),
\label{3DFlo}
\end{alignat}
where $A,B,C$  are restricted to $A,B,C=1,2,3$.
What is important here is that any two regions among 1,2,3, and 4 share not only lines but also finite areas around generic contact point
of all the regions 1,2,3, and 4. Then, the flow is kept unchanged under the deformation of the regions,
as shown in Appendix \ref{s:topo}.
In this sense, we claim that the flow defined above is manifestly topological.

So far we have defined the winding number Eq. (\ref{3DWinNum}) and the flow Eq. (\ref{3DFlo}) in three dimensions. 
Next, we have to consider the relationship between $w_3$ and $f_3$.
The projector $\Pi_x$ is divided into 
\begin{alignat}1
\Pi_x=\sum_{i,j=\pm}\Pi_{+ij},
\end{alignat}
where $\Pi_{+++}$ is the projector onto $j_x\ge0$, $j_y\ge0$, and $j_z\ge0$, 
$\Pi_{++-}$ is the projector onto $j_x\ge0$, $j_y\ge0$, but $j_z<0$, and so on.
Then, we obtain
\begin{alignat}1
w_3(U,\Pi_x,\Pi_y,\Pi_z)&=\sum_{i,j,k,l,m,n=\pm} w_3(U,\Pi_{+ij},\Pi_{k+l},\Pi_{mn+}).
\end{alignat}
In this summation, contributions are three kinds, $\pm f_3/2$ and zero.
To be concrete, let us assign the vector $v_{ijk}=(i,j,k)^T$ for $\Pi_{ijk}$, where $\pm$ in $\Pi_{ijk}$ mean $\pm1$ in $(i,j,k)$, 
and calculate the  determinant $\det (v_{+ij},v_{k+l},v_{mn+})$. If it vanishes, corresponding $w_3$ vanishes. 
For nonzero determinant, let us define the sign of the determinant $s$. 
Then, we have $w_3(\Pi_{+ij},\Pi_{k+l},\Pi_{mn+})=
s\, f_3/2$. 
Exceptions are the cases where three regions spanned by three projectors $\Pi_{+ij},\Pi_{k+l}$, and $\Pi_{mn+}$ separate
the remaining region into two disconnected regions such as $\Pi_{++-},\Pi_{-++}$, and $\Pi_{+-+}$. These are vanishing, even though
the determinants are finite.
Thus, in the summation above, there are $15$ positive-sign terms  and $3$ negative-sign terms which give finite contributions.  
We finally  reach
\begin{alignat}1
w_3(U,\Pi_x,\Pi_y,\Pi_z)&
=6f_3(U,\Pi_1,\Pi_2,\Pi_3),
\end{alignat}
from which it follows 
\begin{alignat}1
{\cal F}_3(U)={\cal W}_3(U).
\label{W3F3Cor}
\end{alignat}
\cin{As in the one-dimensional case of Eq. (\ref{WinNum_1}), it turns out  that the winding number in three dimensions ${\cal W}_3$ 
can  be represented by the flow ${\cal F}_3$ defined on the real lattice space.}
Note that the flow Eq. (\ref{3DFlo}) is well defined in the absence of translational symmetry.
As in the case of one dimension, for numerical computations using finite-size systems, the flow in three dimensions also vanishes trivially. 
Nevertheless, as discussed in Sec. \ref{s:onedimf}, the truncation scheme enables to obtain an approximate winding number.
Namely,
\begin{alignat}1
{\cal F}_3(U)&
=-2\pi i\Tr \left(\epsilon^{ABC}\Pi^{(L)}U^\dagger \Pi_A U\Pi_B U^\dagger \Pi_C U\right)
\label{3DFloTru}
\end{alignat}
is useful for numerical computations, where for simplicity, we assume the truncation projector $\Pi^{(L)}$ as
\begin{alignat}1
\Pi^{(L)}_{ij}=\left\{\begin{array}{ll}\delta_{ij} & (-L\le i_x,i_y,i_z\le L-1)\\0&(\mbox{otherwise})\end{array}\right.
\end{alignat}
for a finite lattice system,  $-N\le i_x,i_y,i_z\le N-1$.

\begin{figure}[h]
\begin{center}
\begin{tabular}{c}
\includegraphics[width=1.\linewidth]{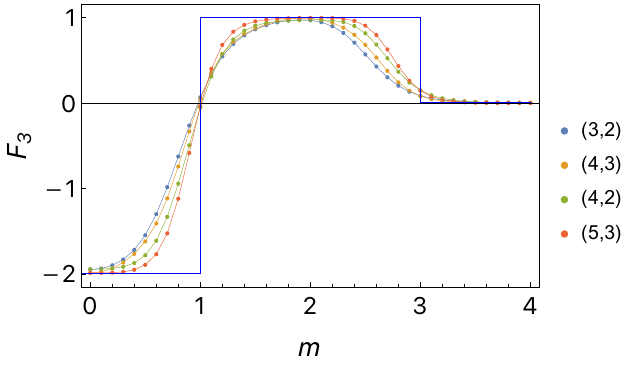}
\end{tabular}
\caption{
The flow ${\cal F}_3$ for $(N,L)=$ $(3,2)$, $(4,3)$, $(4,2)$, and $(5,2)$ systems.
The blue straight lines show the exact winding numbers 
${\cal W}_3=-2,1,0$ for $|m|<1$,  $1<|m|<3$, and  $3<|m|$, respectively,
where we have set $t=b=1$.
}
\label{f:f3}
\end{center}
\end{figure}

\subsection{Application to the Wilson-Dirac model}

Recently, Shiozaki examined \cin{the Dirac operator on the lattice} for
his discrete formula of the winding number \cite{shiozaki2024discrete}. 
Let us compute the flow of  the same  model to check the validity of Eq. \cin{(\ref{W3F3Cor})}.
\cin{To this end, let us start with the Wilson-Dirac model, a typical model of a topological insulator with chiral symmetry in three dimensions, described by
the following Hamiltonian represented by the wave vector
\begin{alignat}1
H&=t\gamma^\mu\sin k_\mu+\gamma^4(m+b\sum_\mu\cos k_\mu).
\end{alignat}
For the $\gamma$ matrices defined by 
$\gamma^\mu=\sigma^1\otimes\sigma^\mu$ for $\mu=1,2,3$ and $\gamma^4=\sigma^2\otimes\sigma^0$, where $\sigma^0$ stands for
the unit matrix, 
it turns out that this model has time-reversal symmetry described by $T=K\sigma^1\otimes i\sigma^2$, where 
$K$ denotes the complex conjugation,  
as well as  chiral symmetry described by $C=\sigma^3\otimes\sigma^0$, and hence, the model belongs to class DIII
\cite{Zirnbauer:1996wi,Altland:1997aa,Schnyder:2008aa,Kitaev:2009fk}.
Thus, the topological property of the half-filled ground state for this model  is 
specified basically by the winding number ${\cal W}_3$ in Eq. (\ref{WinNum_3}). 
With the choice of the $\gamma$ matrices above, the Hamiltonian is represented as}
\begin{alignat}1
\cin{H=\left(\begin{array}{cc}&D\\D^\dagger&\end{array}\right).}
\label{WilDirMod}
\end{alignat}
Here, 
the so-called Wilson-Dirac operator $D$ is defined by
\begin{alignat}1
D&=t\sigma^\mu\sin k_\mu-i \cin{\sigma^0}(m+b \sum_\mu\cos k_\mu)
\nonumber\\
&=\frac{t}{2i}\sigma^\mu(\delta_\mu-\delta_\mu^*)-i\cin{\sigma^0}\left[m+\frac{b}{2}\sum_\mu(\delta_\mu+\delta_\mu^*)\right].
\label{WilDir}
\end{alignat} 
The second line above is the operator represented on the real lattice denoted by
the forward and backward shift operators to the $\mu$ direction,
$\delta_\mu f_{j}=f_{j+\hat\mu}$ and $\delta_\mu^* f_{j}=f_{j-\hat\mu}$, where \cin{$j$ stands for a lattice site} and 
$\hat \mu$ is the unit vector toward $\mu$ direction.

\cin{It is known that there appear three phases with winding number 
$-2$, 1, and 0, depending on the parameters of the model, obtained by the direct computation of the winding number.
Comparing this exact result, let us check the validity of the three-dimensional flow derived in Sec. \ref{s:3Dflow}.}
In Fig. \ref{f:f3}, we show numerical results of the flow \cin{${\cal F}_3$  calculated for $D$
in Eq. (\ref{WilDir}) represented in the real lattice space.
Here, we have firstly unitarized the operator $D$ by the use of the singular value decomposition, as in Eq. (\ref{UniTar}),
and next computed the three-dimensional flow ${\cal F}_3$ given by Eq. (\ref{3DFlo}).}
Contrary to the case of one dimension, the number of lattice sites to each direction are very limited.
\cin{Indeed, the system sizes demonstrated in Fig. \ref{f:f3} are from $6^3$ to $10^3$.} 
Nevertheless,  it turns out that the exact winding number 
is qualitatively reproduced as a function of $m$.
\cin{In particular, in the middle of each phase apart from phase boundaries, the flow is saturated at the exact winding number.}

\begin{figure}[h]
\begin{center}
\begin{tabular}{c}
\includegraphics[width=0.9\linewidth]{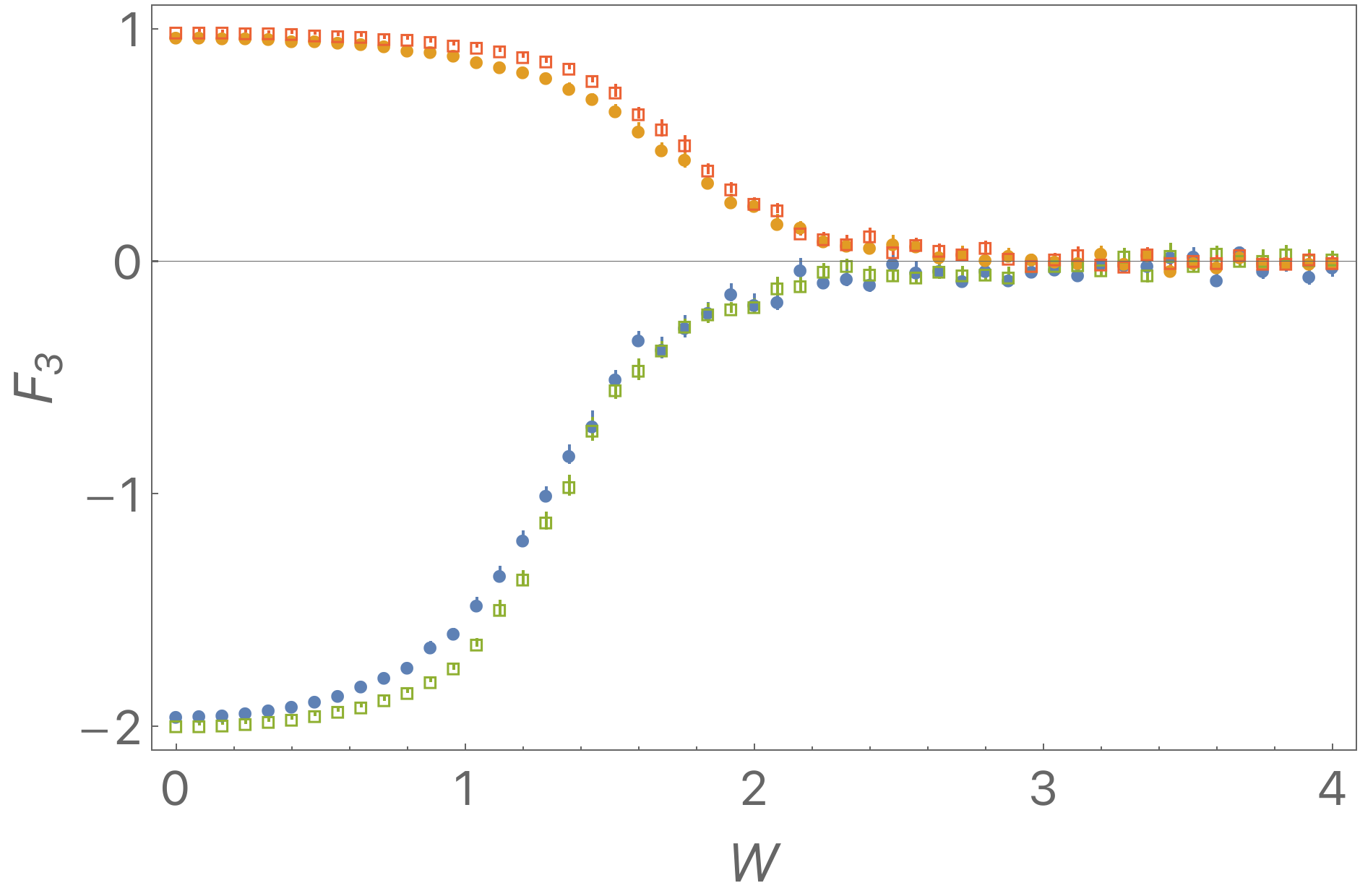}
\end{tabular}
\caption{
\cin{
The flow ${\cal F}_3$ as a function of the disorder strength $W$
for $(N,L)=(3,2)$ (circles) and  $(4,2)$ (squares) systems averaged over 10 ensembles.
Upper two data are for $m=2$, whereas the lower two are for $m=0$.
}
}
\label{f:f3dis}
\end{center}
\end{figure}

\cin{ 
Let us introduce disorder into this model and compute the flow ${\cal F}_3$, instead of the winding number ${\cal W}_3$, 
to see whether topological transitions occur,
since ${\cal W}_3$ is no longer well defined due to broken translational symmetry.
It should be noted here that due to the limited system size, the flow is not very exact, especially near the phase boundary.  
We replace the hopping parameters $t$ and $b$ into site-dependent ones $t_j$ and $b_j$ and set
\begin{alignat}1
t_j=1+\delta t_j,\quad b_j=1+\delta b_j ,
\end{alignat}
where $\delta t_j\in[-W,W]$ and $\delta b_j\in[-W,W]$ are random parameters.
In Fig. \ref{f:f3dis}, the flow ${\cal F}_3$ is shown as a function of the disorder strength $W$.
In both cases $m=0$ and $2$, the transition from topological to trivial phases are observed, but
the analysis of detailed behavior of the transition may need more large systems.
Experimentally, the Wilson-Dirac model can be implemented by the topoelectrical circuits on the hyperbolic lattice.
Indeed, the four-dimensional Wilson-Dirac model has been realized on the hyperbolic $\{8,8\}$ lattice and the second Chern 
number has been observed \cite{Zhang:2023ab}. Freezing one direction of such a circuit, one can obtain the three-dimensional 
circuit corresponding to the Wilson-Dirac model denoted by Eqs. (\ref{WilDirMod}) and (\ref{WilDir}) .
}
\section{Summary and discussions}

The flow of a unitary matrix introduced by Kitaev is a manifestly topological formulation of the winding number 
represented on a real lattice.
Applying to a disordered model with topological transitions, we showed 
that the flow is quite useful for numerical computations with a suitable truncation scheme. 
We also extended  the notion of the flow into three dimensions. 
In such a three-dimensional formulation, our formula reproduces a qualitative feature of the winding number for the 
Wilson-Dirac operator. However, to reveal the quantitative properties, e.g., 
a topological transition of a dirty system, requires numerical ingenuity.

\acknowledgements

This work was supported in part by a Grant-in-Aid for Scientific Research 
	(Grant No. 22K03448) from the Japan Society for the Promotion of Science.
\appendix

\section{Projectors and position operators}

Let $i$ denote a site on a lattice. For the time being, we consider the one-dimensional case.
Let $\theta_i$ be the discrete step function defined by $\theta_i=1$ for $i\ge 0$ and $\theta_i=0$ for $i<0$.
Then, the projection operator can be written as $\Pi_{ij}=\theta_i \delta_{ij}$.
Let $X$ be the position operator defined by $X_{ij}=i\delta_{ij}$.
Let $A_{ij}$ and $B_{ij}$ be matrices which depend on the $i$ and $j$ sites.
Then, we have
\begin{alignat}1
&[\Pi, B]_{ij}=(\theta_i-\theta_j)B_{ij},
\nonumber\\
&[X, B]_{ij}=(i-j)B_{ij}.
\end{alignat}
Thus,
\begin{alignat}1
&(A[\Pi, B])_{ij}=\sum_kA_{ik}(\theta_{k}-\theta_{j})B_{kj},
\nonumber\\
&(A[X, B])_{ij}=\sum_kA_{ik}(k-j)B_{kj}.
\label{PiX}
\end{alignat}
For a system with translational invariance, we assume $A_{ij}=A_{i-j}$. 
Let Tr be the trace over the matrix $A_{ij}$ as well as over the site $i$, i.e.,
$\Tr A=\sum_i\tr A_{ii}$.
Then, 
\begin{alignat}1
\Tr(A[\Pi, B])&=\sum_{i,k}\tr A_{i-k}(\theta_{k}-\theta_i)B_{k-i}
\nonumber\\
&=\sum_{i,k}\tr A_{-k}(\theta_{i+k}-\theta_i)B_{k}
\nonumber\\
&
=\sum_k \tr A_{-k}kB_k=\tr (A[X, B]),
\label{TraInv_1}
\end{alignat}
where we have used $\sum_i (\theta_{i+j}-\theta_{i+k})=j-k$.
Note that the last equation does not depend on $i$ for translational invariance, 
as can be seen from Eq. (\ref{PiX}) in the case of  $i=j$.

Translational invariance also enables to make the Fourier transformation
\begin{alignat}1
A_{j}=\int_{-\pi}^\pi\frac{dq}{2\pi}e^{iqj}A_q.
\end{alignat}
It follows that
\begin{alignat}1
\tr (A[X, B]) =\frac{i}{2\pi}\int_{-\pi}^\pi dq \tr A_q\partial_q B_q.
\label{TraInv_2}
\end{alignat}

The above $\Pi$-$X$ correspondence, Eq. (\ref{TraInv_1}) 
is valid only in the case of the trace of matrices including  a single commutator of $\Pi$ and $X$.
In the three-dimensional case,  the winding number has three commutators of $X_\mu$ and $\Pi_\mu$, 
but they are different directions. 
Therefore, Eq. (\ref{TraInv_1}) can be applied.

\section{The flow represented by zero modes}\label{s:index}

Given a unitary matrix, we can define a Hermitian matrix in doubly extended space.
In this section, we show that the flow has an intimate relationship with the zero modes of such a Hermitian matrix.
It may be a kind of the index theorem.

From a unitary matrix Eq. (\ref{UniTar}),
let us define the Hermitian operator (or the SSH-like Hamiltonian but with the projector) as
\begin{alignat}1
H&=\left(\begin{array}{cc}&\iPi U \\ U^\dagger\iPi&\end{array}\right)
\nonumber\\
&=\left(\begin{array}{cc}\iPi& \\ &\1\end{array}\right)
\left(\begin{array}{cc}&U \\ U^\dagger&\end{array}\right)
\left(\begin{array}{cc}\iPi&\\&\1\end{array}\right).
\label{Ham}
\end{alignat}
For a while, we consider the infinite system in Sec. \ref{s:finite}.
This operator $H$ has chiral symmetry 
\begin{alignat}1
\Gamma H\Gamma^{-1}=-H,\quad 
\Gamma\equiv \left(\begin{array}{cc}\1&\\&-\1\end{array}\right),
\end{alignat}
where $\1$ stands for the identity matrix in the space of $U$.
Note that 
\begin{alignat}1
H^2=\left(\begin{array}{cc}\iPi&\\&U^\dagger\iPi U\end{array}\right).
\end{alignat}
Then, the flow Eq. (\ref{DefFlo}) can be written by
\begin{alignat}1
{\cal F}_1=-\widetilde{\rm Tr}\,\Gamma H^2,
\end{alignat}
where $\widetilde{\rm Tr}$ stands for the trace in the extended space.
Using 
$\widetilde{\rm Tr}\,\Gamma=0$,
we have
\begin{alignat}1
{\cal F}_1=\widetilde{\rm Tr}\, \Gamma(1-H^2).
\end{alignat}
The operator $H$ has eigenvalues $\pm1$ and $0$.
Let $\varphi_{n}$ ($n=1,2,\dots,$) be the wave function of $H$ with eigenvalues $1$.
Then, 
because of chiral symmetry, the wave functions with eigenvalue $-1$ denoted by $\varphi_{-n}$ can be given by
$\varphi_{-n}=\Gamma \varphi_{n}$ ($n=1,2,\dots$).
Namely, the $\pm1$ energy states are always paired.
Let $\varphi_{0m}$ ($m=1,2,\dots$) be the wave function of $H$ with eigenvalue $0$.
Such zero-energy states can be chosen as the eigenstates of $\Gamma$. 
Therefore, the zero modes are classified as $\varphi_{0m}^\pm$, where 
$\Gamma\varphi_{0m}^\pm=\pm\varphi_{0m}^\pm$.
To be concrete, they are solutions of 
\begin{alignat}1
U^\dagger\iPi\varphi_{0m}^+=0,\quad \iPi U\varphi_{0m}^-=0.
\end{alignat}
Using the eigenfunctions, the flow can be expressed as
\begin{alignat}1
{\cal F}_1&=\sum_{m}\varphi^\dagger_{0m}\Gamma\varphi_{0m}
=\#(+)-\#(-).
\end{alignat}
Namely, the flow is just the difference of the numbers of the zero modes of $H$.
However, note that this is just a formal result, since $\#(\pm)$ are,  respectively, infinite.
Therefore, some regularization should be needed.
Moreover, the flow vanishes for finite systems.

For practical purposes, the truncation scheme mentioned in the text is also useful:
\begin{alignat}1
{\cal F}_1&=\widetilde{\rm Tr}\, \Gamma\widetilde\Pi^{(L)}(1-H^2)
\nonumber\\
&=\sum_m\varphi_{0m}\Gamma\widetilde\Pi^{(L)}\varphi_{0m}
\nonumber\\
&=\sum_m\varphi_{0m}^+\widetilde\Pi^{(L)}\varphi_{0m}^+-\sum_m\varphi_{0m}^-\widetilde\Pi^{(L)}\varphi_{0m}^-,
\label{FloNonUni}
\end{alignat}
where $\widetilde\Pi^{(L)}\equiv {\rm diag}(\Pi^{(L)},\Pi^{(L)})$ stands for the projector $\Pi^{(L)}$ 
extended to the doubled space of $H$.
This formula is valid for finite systems. 

\section{Flow as a topological invariance}\label{s:topo}

We show that the flow
is manifestly topological.
\cin{To begin with, let us start in the case of one dimension to fix our notations.
In this Appendix \ref{s:topo}, a unitary matrix $U$ is denoted as $U_{i_1i_2}$, where $i_a$ specifies the number of the 
unit cell as well as the number of internal degrees of freedom inside the unit cell such as $i_a=(i,s)$.
Then, according to Kitaev, let us define the current
\begin{alignat}1
f_{i_1i_2}&\equiv U^\dagger_{i_1i_2}U_{i_2i_1}-U^\dagger_{i_2i_1}U_{i_1i_2}
= \epsilon^{ab}U^\dagger_{i_ai_b}U_{i_bi_a},
\end{alignat}
where $a,b=1,2$ are implicitly summed  in the last equality.
Note that 
\begin{alignat}1
f_{i_1i_2}=-f_{i_2i_1}.
\label{AntSym}
\end{alignat} 
The current is conserved: For a fixed $i_1$, we have
\begin{alignat}1
\sum_{i_2}f_{i_1i_2}&=\sum_{i_2}\left(U^\dagger_{i_1i_2}U_{i_2i_1}-U^\dagger_{i_2i_1}U_{i_1i_2}\right)
\nonumber\\
&=1-1=0,
\label{FloCon}
\end{alignat}
where the sum over $i_2=(i,s)$ means the sum over $i$ and $s$, and 
we have used the fact that $U$ is unitary.
The flow in Eqs. (\ref{Flo}) or (\ref{DefFlo}) is given by
\begin{alignat}1
{\cal F}_1(U)=\sum_{i_1\in1}\sum_{i_2\in2}f_{i_1i_2}.
\end{alignat}
Now let us pick up a specific site $i_0\in 1$, and define the new region $1'$ excluding $i_0$, i.e., 
$1=1'+i_0$. Then, the above flow can be written as
\begin{alignat}1
{\cal F}_1(U)=\sum_{i_1\in1'}\sum_{i_2\in2}f_{i_1i_2}+\sum_{i_2\in2}f_{i_0i_2}.
\end{alignat}
If one reassigns $i_0$ to region 2, the flow changes into
\begin{alignat}1
{\cal F}_1'(U)=\sum_{i_1\in1'}\sum_{i_2\in2}f_{i_1i_2}+\sum_{i_1\in1'}f_{i_1i_0}.
\end{alignat}
The difference is 
\begin{alignat}1
{\cal F}_1-{\cal F}_1'&=\sum_{i_2\in2}f_{i_0i_2}-\sum_{i_1\in1'}f_{i_1i_0}
\nonumber\\
&=\sum_{i_2\in2}f_{i_0i_2}+\sum_{i_1\in1'}f_{i_0i_1}+f_{i_0i_0}
\nonumber\\
&=\sum_{i_2}f_{i_0i_2}=0,
\end{alignat}
where we have used Eqs. (\ref{AntSym}) and (\ref{FloCon}).
Thus, the flow is invariant under the reassignment of a site into another region. 

In the three-dimensional case, the invariance of flow can also be shown in parallel with the one-dimensional case.}
Let us define a current in three dimensions
\begin{alignat}1
f_{i_1i_2i_3i_4}=\epsilon^{abcd}U^\dagger_{i_ai_b}U_{i_bi_c}U^\dagger_{i_ci_d}U_{i_di_a},
\end{alignat}
where $i_a$ ($a=1,2,3,4$) specifies  a site in three dimensions.
By definition, $f_{i_1i_2i_3i_4}$ is antisymmetric in all four indices $i_a$. 
We first show that the current is conserved at each site:
\begin{alignat}1
&\sum_{i_4}f_{i_1i_2i_3i_4}
\nonumber\\
&=\epsilon^{abc}\Big[
U^\dagger_{i_ai_b}U_{i_bi_c}U^\dagger_{i_ci_4}U_{i_4i_a}
-U^\dagger_{i_ai_b}U_{i_bi_4}U^\dagger_{i_4i_c}U_{i_ci_a}
\nonumber\\
&
\qquad\quad+U^\dagger_{i_ai_4}U_{i_4i_b}U^\dagger_{i_bi_c}U_{i_ci_a}
-U^\dagger_{i_4i_a}U_{i_ai_b}U^\dagger_{i_bi_c}U_{i_ci_4}
\Big]
\nonumber\\
&=\epsilon^{abc}\Big[
U^\dagger_{i_ai_b}U_{i_bi_a}\delta_{i_ai_c}-U^\dagger_{i_ai_b}U_{i_bi_a}\delta_{i_bi_c}
\nonumber\\
&\qquad\quad+U^\dagger_{i_ai_c}U_{i_ci_a}\delta_{i_ai_b}-
U^\dagger_{i_bi_a}U_{i_ai_b}\delta_{i_ai_c}\Big]
\nonumber\\
&=\epsilon^{abc}\delta_{i_ai_c}\Big[
U^\dagger_{i_ai_b}U_{i_bi_a}+U^\dagger_{i_bi_a}U_{i_ai_b}
\nonumber\\
&\qquad\qquad\quad-U^\dagger_{i_ai_b}U_{i_bi_a}-U^\dagger_{i_bi_a}U_{i_ai_b}\Big]
\nonumber\\
&=0,
\label{ConCur}
\end{alignat}
where $a,b,c$ are restricted to 1,2,3, \cin{and
repeated $i_4$ in the unitary matrices are implicitly summed.}
Using the current, we can write the flow ${\cal F}_3$ such that
\begin{alignat}1
{\cal F}_3(U)=2\pi i\sum_{i_1\in1}\sum_{i_2\in2}\sum_{i_3\in3}\sum_{i_4\in4}f_{i_1i_2i_3i_4},
\label{Flo3}
\end{alignat}
where $1,2,3,4$ stand for the regions in the three-dimensional lattice introduce above Eq. (\ref{DefFlo3}).

Next, let us show that the flow Eq. (\ref{Flo3}) is topological, since the flow is invariant even if a site in a region 
is assigned  to another region, implying that the flow does not depend on the detailed shapes of regions 1,2,3, and 4.
Let  $i_0\in 1$ be a site in region 1,  and let $1'$  be the set of sites in region 1 except for $i_0$, i.e., $1=i_0+1'$.
Then, 
\begin{alignat}1
{\cal F}_3=2\pi i f_{1234}\equiv 2\pi i\left(f_{1'234}+f_{i_0234}\right),
\end{alignat}
where $f_{1ijk}=\sum_{\l\in 1}f_{lijk}$, and so on. Let us assign $i_0$ in region 2.  Then, the flow becomes
 \begin{alignat}1
{\cal F}_3'=2\pi i\left(f_{1'234}+f_{1'i_034}\right).
\end{alignat}
The difference is
\begin{alignat}1
{\cal F}_3-{\cal F}_3'&=2\pi i\left(f_{i_0234}-f_{1'i_034}\right)
\nonumber\\
&=2\pi i\left(f_{i_0234}+f_{i_01'34}\right).
\label{F3Def}
\end{alignat}
On the other hand, the conservation of the current Eq. (\ref{ConCur}) can be written as 
\begin{alignat}1
f_{1'jkl}+f_{i_0jkl}+f_{2jkl}+f_{3jkl}+f_{4jkl}=0,
\end{alignat}
for fixed $j,k,l$, when $i_0\in1$.
Sum over $j\in3$ and $k\in4$ in the above conservation law yields
\begin{alignat}1
0&=f_{1'34l}+f_{i_034l}+f_{234l}+f_{334l}+f_{434l}
\nonumber\\
&=f_{1'34l}+f_{i_034l}+f_{234l}.
\end{alignat}
Setting $l=i_0$, we have
\begin{alignat}1
0=f_{1'34i_0}+f_{i_034i_0}+f_{234i_0}=f_{1'34i_0}+f_{234i_0}.
\end{alignat}
It follows from Eq. (\ref{F3Def}) that ${\cal F}_3={\cal F}_3'$.


\end{document}